\documentclass[nofootinbib]{revtex4}
\usepackage{graphicx}
\usepackage{amsmath,amssymb,epsfig}

\allowdisplaybreaks

\renewcommand{\vec}[1]{\mathbf{#1}}
\let\Re=\relax
\DeclareMathOperator{\Re}{Re}

\begin{document}

\title{Image recovery with the solar gravitational lens}
\author{Viktor T. Toth$^1$, Slava G. Turyshev$^2$}

\affiliation{\vskip 3pt
$^1$Ottawa, Ontario K1N 9H5, Canada}

\affiliation{\vskip 3pt
$^2$Jet Propulsion Laboratory, California Institute of Technology,\\
4800 Oak Grove Drive, Pasadena, CA 91109-0899, USA}

\date{\today}

\begin{abstract}
We report on the initial results obtained with an image convolution/deconvolution computer code that we developed and used to study the image formation capabilities of the solar gravitational lens (SGL). Although the SGL of a spherical Sun creates a greatly blurred image, knowledge of the SGL's point-spread function (PSF) makes it possible to reconstruct the original image and remove the blur by way of deconvolution. We discuss the deconvolution process, which can be implemented either with direct matrix inversion or with the Fourier quotient method. We observe that the process introduces a ``penalty'' in the form of a reduction in the signal-to-noise ratio (SNR) of a recovered image, compared to the SNR at which the blurred image data is collected. We estimate the magnitude of this penalty using an analytical approach and confirm the results with a series of numerical simulations. We find that the penalty is substantially reduced when the spacing between image samples is large compared to the telescope aperture.  The penalty can be further reduced with suitable noise filtering, which can yield ${\cal O}(10)$ or better improvement for low-quality imaging data. Our results confirm that it is possible to use the SGL for imaging purposes. We offer insights on the data collection and image processing strategies that could yield a detailed image of an exoplanet within image data collection times that are consistent with the duration of a realistic space mission.
\end{abstract}


\maketitle

\section{Introduction}

According to the general theory of relativity \cite{Einstein:1915}, large, gravitating objects such as the Sun bend rays of light. The resulting solar gravitational lens (SGL) may be used in as an instrument provided by Nature: a part of an immensely powerful telescope with very large light amplification and significant angular resolution capabilities \cite{Turyshev-Toth:2017}.

The focal region of the SGL begins beyond $\sim 550$ astronomical units (AU) from the Sun. This is almost four times the distance to our most distant spacecraft to date, Voyager 1, which is over 150 AU as of late 2020. The SGL is an imperfect lens, which suffers from spherical aberration and astigmatism \cite{Turyshev-Toth:2020-extend}. Furthermore, though the SGL's light amplification is tremendous, any signal from a faint, distant source is overwhelmed by light from the Sun itself, as well as from the solar corona. These challenges must be addressed if the SGL is to be considered as a practical ``instrument'' for high-resolution observations of distant, extrasolar targets.

In previous papers \cite{Turyshev-Toth:2017,Turyshev-Toth:2018,Turyshev-Toth:2018-plasma,Turyshev-Toth:2018-grav-shadow,SGL2018f,Turyshev-Toth:2019,Turyshev-Toth:2019-extend,Turyshev-Toth:2019-blur,Turyshev-Toth:2019-image,Turyshev-Toth:2020-extend}, we developed a wave-theoretical description of the SGL starting from the first principles of Maxwell's theory of electromagnetism on the curved background metric of the solar gravitational field. We accounted for the monopole gravitational field of the Sun and contributions (negligible, as it turned out, at optical or near-IR wavelengths) from the charged medium of the solar corona. Our work led to establishing the SGL's optical properties and, in particular, its point-spread function (PSF), which is used  to characterize the imaging process with the lens, especially in the context of a deep space mission \cite{Turyshev-etal:2018,Turyshev-etal:2020}.

This PSF can now be used to directly simulate the imaging data that is produced by the SGL's action on the imaging signal received from a  distant source  (convolution), and also the reconstruction (deconvolution) of that image. These steps can be implemented in computer code, providing valuable insight into the nature of the SGL's PSF and the requirements and limitations of any deconvolution process.

In the remainder of this paper, we first introduce the SGL's PSF in Section~\ref{sec:SGL} and discuss its properties related to image formation. We describe image deconvolution using the method of Fourier quotients. The computationally more demanding method of direct deconvolution is also addressed and used to develop an assessment of the resulting change in the signal-to-noise ratio (SNR) that we call the ``deconvolution penalty''. We discuss these results in Section~\ref{sec:RESULT} and present our conclusions and future plans in Section~\ref{sec:CONCL}.

\section{Properties of the solar gravitational lens}
\label{sec:SGL}

It has been known since 1915 \cite{Einstein:1915} that massive objects deflect rays of light, with an angle of deflection $\delta= 2r_g/b$, where $r_g=2GM/c^2$ is the object's Schwarzschild radius,  $G$ is the Newton's constant of gravitation, $M$ is the mass of the object, $c$ is the speed of light (yielding for the Sun $r_g=2.95$ km) and $b$ is the light ray's impact parameter with respect to the center-of-mass of the massive object. As light rays that pass by the Sun are bent ``inward'', rays from opposite sides of the Sun are expected to eventually meet some distance away from the Sun \cite{Turyshev-Toth:2017}.

\subsection{Imaging geometry with the SGL and its PSF}
\label{sec:PSF}

\begin{figure*}
\includegraphics{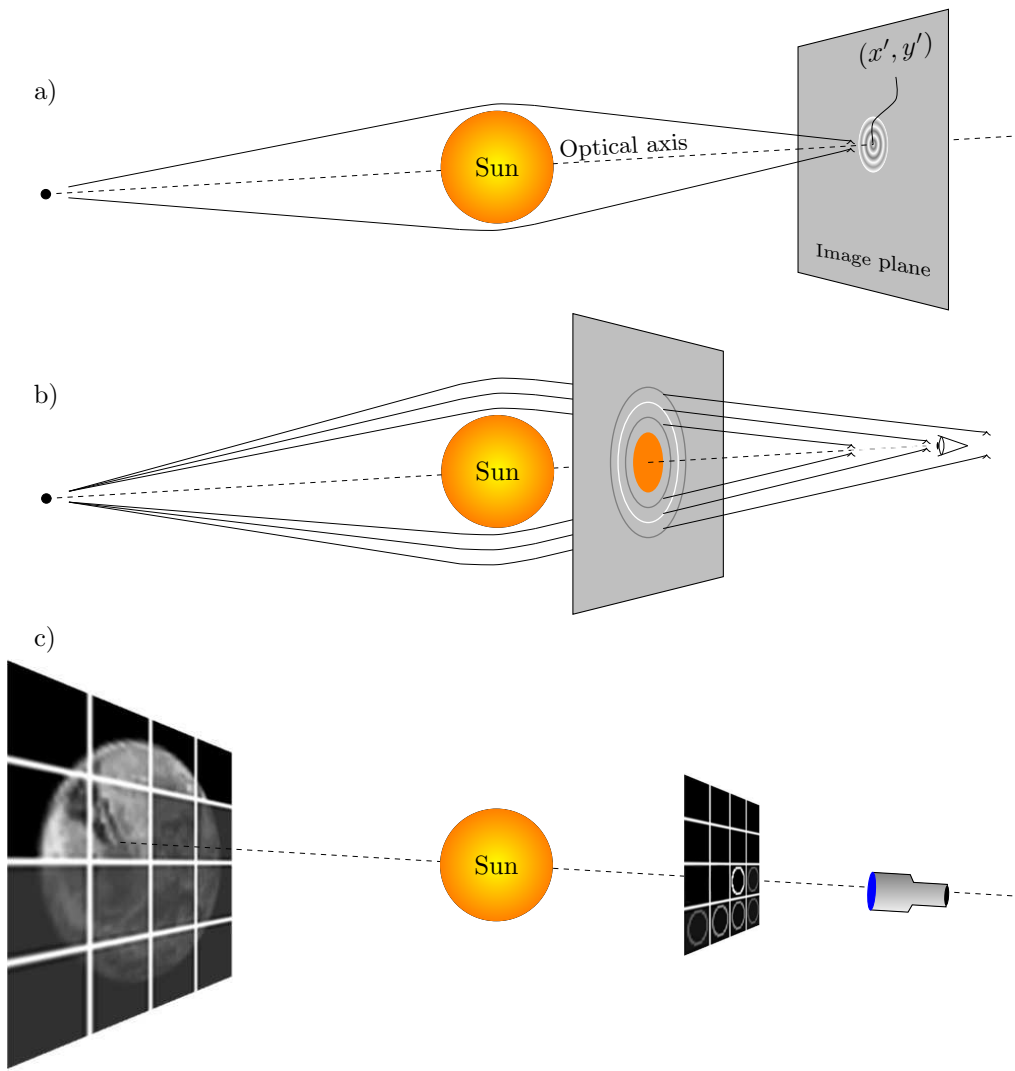}
\caption{\label{fig:SGL1}The SGL and the effect of its PSF on light from a point source: a) the Airy-pattern that the SGL projects onto the image plane; b) the Einstein-ring (white) seen by an observer located at the focal region, looking back at the Sun (two additional Einstein-rings, which would be seen by observers nearer the Sun or farther from the Sun, are shown in gray); c) Image sampling as a telescope scans the image plane, while measuring the varying intensity of the Einstein-ring of an extended source (which may be modeled as a multitude of point sources, each of which contributes to various portions of the observed Einstein-ring) as seen from different vantage points, mapping these intensities into the corresponding image pixels.}
\end{figure*}

Considering the geometry of the image formation process with the SGL, we observe that parallel rays of light coming from infinity and just grazing the Sun converge at a point that is located at the distance given by $z=R_\odot/\sin\delta$ from the Sun, where $\delta$ is the angle of deflection and $R_\odot= 6.96\times 10^8$~m is the solar radius. For grazing rays $\delta=2r_g/R_\odot\sim 1.76''$, yeilding $z\sim 550$~AU (see Fig.~\ref{fig:SGL1}a).

However, unlike a well-constructed optical thin lens, the SGL does not focus light from a distant point source to a point. Rays of light with larger impact parameters, $b$, reach the ``optical axis'' (the imaginary line connecting the distant point source with the center of the Sun) at greater and greater distances. Thus, instead of a focal point the SGL forms a focal half-line (Fig.~{\ref{fig:SGL1}}b).

Light from a point source would appear, to an observer on the focal half-line and looking back at the Sun, as a circle of light around the Sun: the Einstein-ring. The farther the observer is located from the Sun, the larger the Einstein-ring appears relative to the Sun \cite{Turyshev-Toth:2019-image,Turyshev-Toth:2020-extend}. If the observer moves away from the focal line, the Einstein-ring from a point source would rapidly break into two arcs; further away from the focal line, the two arcs each collapse into two spots of light. Much further away (at distances comparable to the solar radius) the two spots would become asymmetric in appearance, one eventually vanishing behind the Sun, while the other transitioning into the unlensed image of the distant source.

Back at the focal line, at a specific, given distance from the Sun, light from a distant point source is spread out in a pattern. In addition to the light that arrives at the focal half-line, there will be light spread around it, from rays of light with impact parameters that are either smaller or greater than the impact parameter corresponding to the observer's distance from the Sun. This is the pattern that is determined by the SGL's point-spread function, or PSF.

\begin{figure*}
\includegraphics{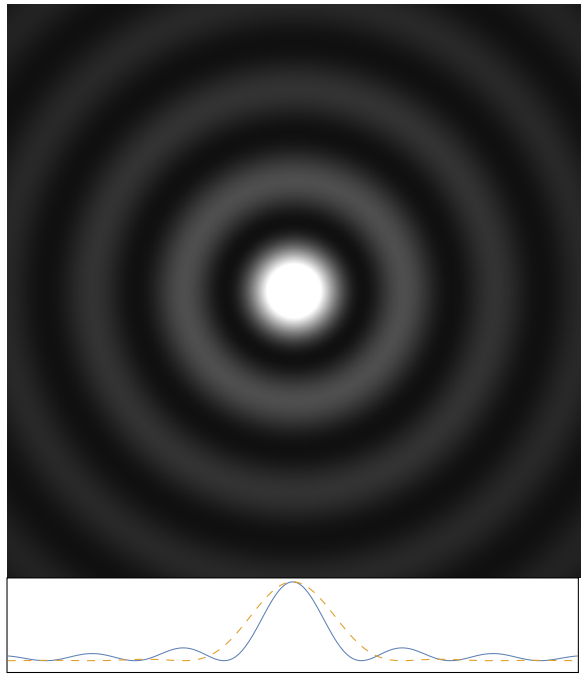}
\caption{\label{fig:PSF}The PSF of the SGL. At typical wavelengths, in the image plane, the spatial frequency of this pattern is on the scale of a few centimeters; the average amplitude is proportional to the inverse of the radial distance from the center of the pattern. Bottom panel shows the relative height of the first few peaks in dimensionless units. For comparison with the SGL PSF, $\propto J^2_0(\alpha\rho)$, a typical thin lens PSF (dashed line, $\propto J^2_1(\alpha\rho)/\rho^2$) is also shown.}
\end{figure*}

For a point source of light, a generic PSF captures how light from that source is deposited in an image plane (Fig.~\ref{fig:PSF}). (This is an image plane into which the SGL projects an image, and must not be confused with the image plane that would be observed by an imaging telescope looking back at the Sun.) In the most general case, the PSF depends on two vector-valued parameters: the location $\vec{x}'$ of the point source in the source plane (or equivalently, the image plane position $\beta\vec{x}'$ of the intersection of the optical axis and the image plane, with the coordinate scaling factor $\beta=-z/z_0$ constructed using the distance $z_0$ between the image source and the Sun \cite{Turyshev-Toth:2019-extend}) and the position $\vec{x}$ in the image plane where light intensity is measured. The generic PSF, which describes light received at $\vec{x}$ in the image plane due to a point source whose optical axis intersects the image plane at $\vec{x}'$ can be written in the form ${\tt PSF}(\vec{x},\vec{x}')$.

The PSF of the SGL can be derived in a variety of ways (see \cite{Turyshev-Toth:2017} and references therein). We presented a particularly rigorous derivation in \cite{Turyshev-Toth:2019}, where we studied the combined effects of solar gravity and the electrically charged solar corona on rays of light grazing the Sun, starting with Maxwell's field equations on the curved background of the solar gravitational field. Assuming a spherically symmetric Sun, we obtained a PSF in the form
\begin{equation}
{\tt PSF}(\rho)= J^2_0\left(\alpha\rho\right),
\label{eq:PSF}
\end{equation}
where $\rho=|\vec{x}+\beta\vec{x}'|$ and $\alpha=(2\pi/\lambda)\sqrt{2r_g/z}$ is a quantity constructed from the observational wavelength $\lambda$, the Sun's Schwarzschild radius $r_g$, and the distance $z$ from the center of the Sun to the image plane. The function $J_0(z)$ is the $0$-th Bessel function of the first kind. This PSF is depicted in Fig.~\ref{fig:PSF}. Note that the amplitude of (\ref{fig:PSF}) decreases slowly, as $1/\rho$. This is a result of the spherical aberration of the SGL, in contrast with a thin lens with no spherical aberration and a PSF that has an amplitude that is proportional to $1/\rho^3$ \cite{Turyshev-Toth:2020-extend}.

\subsection{The effective PSF for a finite aperture telescope}

At $z=550$~AU from the Sun, an optical telescope that is capable of resolving the solar disk and also able to accommodate a coronagraph to block out sunlight must have a meter-class aperture, $d\gtrsim 1$~m. This is much larger than the spatial periodicity of the SGL PSF in the image plane at optical or near-IR wavelengths, which, based on (\ref{eq:PSF}), is measured in centimeters.

Although such a telescope is capable of forming a resolved image of the Einstein-ring that appears around the Sun, this is important only insofar as it allows us to block out the glare of the Sun and reduce the amount of coronal light using an appropriately constructed coronagraph or similar technique. Ultimately, what is of interest to us is the total amount of light received from the Einstein-ring at a given telescope location (see Fig.~\ref{fig:SGL1}c). This corresponds to the amount of light from the Einstein-ring that is collected by the telescope aperture. To estimate this quantity, it is necessary to average the PSF over the telescope aperture. In \cite{Turyshev-Toth:2020-extend}, we obtained a very accurate approximation of this averaged SGL PSF in the form:
\begin{equation}
\overline{\tt PSF}(\rho)=\frac{4}{\pi\alpha d}\mu(\rho),
\label{eq:thePSF}
\end{equation}
where
\begin{equation}
\mu(\rho)=
\Re\left(\dfrac{2}{\pi}{\tt E}\Big[\arcsin\Big(\dfrac{d}{2\rho}\Big),\Big(\dfrac{2\rho}{d}\Big)^2\Big]\right),
\label{eq:mu}
\end{equation}
where ${\tt E}[\alpha,z]$ is the incomplete elliptic integral \cite{Abramovitz-Stegun:1965}. (We note that for $\rho\le\tfrac{1}{2}d$,
$\mu(\rho)={\tt E}[(2\rho/d)^2]$ where ${\tt E}[z]$ is the complete elliptic integral.)

The PSF characterized by Eq.~(\ref{eq:mu}) falls off very slowly. For ${2\rho}/{d}\gg 1$, $\mu(\rho)=d/4\rho$ is a close approximation \cite{Turyshev-Toth:2020-extend}. Thus, light from a point source is spread over a large area. This presents special challenges when it comes to image formation and image deconvolution, as we shall see in the next section. On the other hand, we note that this averaged form lost any dependence on the wavelength $\lambda$, thus it is not necessary to account for any wavelength-dependent behavior.

\subsection{Image formation and reconstruction}

Given an original image in a source plane characterized by intensities $O(\vec{x}')$, an image in the image plane is formed by a lens characterized by a PSF given by ${\tt PSF}(\vec{x},\vec{x}')$ by the convolution
\begin{align}
I(\vec{x})=\iint d^2\vec{x}'{\tt PSF}(\vec{x},\vec{x}')O(\vec{x}').
\label{eq:convolve}
\end{align}

To understand the geometry of image formation, consider that light from an Earth-sized exoplanet, located at $\sim 30$~pc from the solar system, is projected onto an approximately $1300\times 1300$~m area at a distance of 650~AU from the Sun (see Fig.~\ref{fig:SGL1}). An observer, equipped with a suitable telescope and coronagraph, positioned in this region and looking back at the Sun, would see an Einstein-ring around the Sun. The intensity of the Einstein-ring changes as the observer changes location, and the Einstein-ring is dominated by light from different regions of the distant target. Outside the projected image of the exoplanet, the observer would see the Einstein-ring break up into arcs.

It is obviously not practical to capture such a large image all at once, as it would require a square kilometer size instrument. Instead, we envision an observer moving in the image plane and sampling the light field at different locations, essentially scanning it one large (square meter scale) picture element (pixel) at a time. The sole observable at every measurement location is the total intensity of light received from the exoplanet, i.e., the integrated brightness of the Einstein-ring seen around the Sun. The observer may use an imaging telescope to look in the direction of the Sun, but the purpose of this is to separate sunlight and light from the exoplanet, perhaps by blocking out sunlight using a coronagraph. Ultimately, while details of the Einstein-ring may contain additional useful information, only its total brightness is required in in principle for image reconstruction.

The projected image of the exoplanet is blurred because of the spherical aberration of the SGL. This can be modeled by convolving a source image with the PSF of the SGL. Image reconstruction therefore requires deconvolution: inverting this convolution to recover $O(\vec{x}')$ from the values of $I(\vec{x})$ measured in the image plane.

We investigate two distinct but related approaches for deconvolution: direct deconvolution and Fourier deconvolution.

A key concern is that deconvolution has a disproportionate effect on noise. That is to say, if the convolved image, $I(\vec{x})$, is observed in the presence of noise, the ratio of signal-to-noise will increase as the original image $O(\vec{x}')$ is recovered from the convolved image. This noise amplification can be investigated methodically in the context of direct deconvolution, which is what we study first.

\subsection{Direct deconvolution}
\label{sec:DIRECT}

To recover $O(\vec{x}')$ from $I(\vec{x})$, we begin by discretizing the integral (\ref{eq:convolve}). Dividing the source area into a uniformly distributed grid of $N$ equal-size segments (e.g., a square grid), centered on locations $\vec{x}_i'~(i\in[1,N])$ allows us to rewrite this integral as a sum:
\begin{equation}
I(\vec{x})=\sum_{i=1}^N{\tt PSF}(\vec{x},\vec{x}_i')O(\vec{x}_i').
\end{equation}
Assuming that we sample the image plane at $N$ locations $\vec{x}_j~(j\in[1,N])$, we can define the {\em convolution matrix} as follows (with summation over repeat indices implied):
\begin{equation}
I_j=\sum_{i=1}^N{\tt PSF}(\vec{x}_j,\vec{x}_i')O(\vec{x}_i')=C_{ij}O_i,
\label{eq:IfromJ}
\end{equation}
where
$I_j=I(\vec{x}_j)$ is the brightness of the signal received at the $j$-th pixel on the image plane,  $O_i=O(\vec{x}_i')$ is the brightness at the $i$-th pixel in the source plane and $C_{ij}={\tt PSF}(\vec{x}_j,\vec{x}_i')$ is the convolution matrix of $N^2$ elements.

Knowledge of the PSF and the coordinates of $\vec{x}_j$ and $\vec{x}_i'$ yields $C_{ij}$. If the inverse of this square matrix exists, the original image can be recovered by simple matrix inversion:
\begin{equation}
O_i= C_{ij}^{-1}I_j.
\label{eq:inv}
\end{equation}
Of course inverting a large matrix (for a megapixel image, $N=10^6$, the convolution matrix has $N^2=10^{12}$ elements) is computationally costly and numerically unstable; a better approach is to use standard algorithms to solve the linear system of equations (\ref{eq:IfromJ}) for the unknowns $O_j$.

In general, the locations $\vec{x}_j$ and $\vec{x}_i'$ in the image plane need not coincide. In principle, it is even possible to recover the image of a target that lies entirely outside the ``directly imaged'' region corresponding to a sampled image area; or, it is possible to treat the system as overdetermined (i.e., use fewer $O_j$ than the number of $I_i$ measurements available) with a non-square convolution matrix and employ standard optimization algorithms to find a best-fit solution. However, when the point sets $\vec{x}_j$ and $\vec{x}_i'$ do coincide (i.e., when the locations are chosen such that $\vec{x}_i-\vec{x}_i'=0$), the convolution matrix is square, symmetric, and it is dominated by its diagonal \cite{Turyshev-Toth:2020-extend}.

\subsection{Fourier deconvolution}
\label{sec:DECONV}

Direct deconvolution is computationally expensive. In some cases, it is possible to speed up deconvolution very significantly by performing it in Fourier-space. This is possible as a result of the Fourier convolution theorem, according to which, in specific situations, the computationally costly matrix inversion required to compute (\ref{eq:inv}) or, equivalently, explicitly solving the linear system of equations (\ref{eq:IfromJ}) for $O(\vec{x}'_i)$ given known values of $I(\vec{x}_j)$, can be replaced by simple division. This method is sometimes referred to as the method of Fourier quotients.

In our case, the PSF can be written in the form,
\begin{align}
{\tt PSF}(\vec{x},\vec{x}')={\tt PSF}\big(\vec{x}+\beta\vec{x}'\big).
\end{align}
Fourier-transforming $I(\vec{x})$ leads to the following result, known as the Fourier convolution theorem \cite{Goodman:1968}:
\begin{align}
\hat{I}({\bf f})&=\iint d^2{\bf x}e^{-2\pi i{\bf x}\cdot{\bf f}}I({\bf x})\\
&=\iint d^2{\bf x}e^{-2\pi i{\bf x}\cdot{\bf f}}\iint d^2{\bf x}'O({\bf x}'){\tt PSF}({\bf x}+\beta{\bf x}')\nonumber\\
&=\iint d^2{\bf x}'e^{-2\pi i(-\beta{\bf x}')\cdot{\bf f}}O({\bf x}')\iint d^2{\bf x}e^{-2\pi i({\bf x}+\beta{\bf x}')\cdot{\bf f}}{\tt PSF}({\bf x}+\beta{\bf x}')\nonumber\\
&=\widehat{\tt PSF}({\bf f})\iint d^2{\bf x}'e^{-2\pi i{\bf x}'\cdot(-\beta{\bf f})}O({\bf x}') =\widehat{\tt PSF}({\bf f})\hat{O}(-\beta{\bf f}),\nonumber
\end{align}
hence,
\begin{equation}
\hat{O}(-\beta{\bf f})\propto\frac{\hat{I}({\bf f})}{\widehat{\tt PSF}({\bf f})}.\label{eq:deconv}
\end{equation}
This is how, in Fourier-space, deconvolution is reduced to simple division. Used in conjunction with fast Fourier transform (FFT) algorithms, this approach dramatically reduces the computational complexity of the problem of image deconvolution.

It is important to make note of some caveats, however, regarding the limits of applicability of Fourier methods.

First, in this formulation we utilized the fact that the PSF is a function of $\vec{x}+\beta\vec{x}'$, and not individually dependent on $\vec{x}$ and $\vec{x}'$. Such a dependence may exist, for instance, if various points $\vec{x}'$ are imaged at different times, with a time-varying PSF reflecting temporal changes in the orientation, illumination or appearance of the source. In this case, the method of Fourier quotients is not applicable.

Second, the Fourier method necessarily introduces artifacts through the implied assumptions that underlie the finite Fourier transform. Direct convolution transforms light from a finite area in the source plane, assuming no additional light from outside the defined source area. The Fourier method, in contrast, amounts to the assumption that the finite source area is an accurate spectral representation of the entire (infinite) source plane. In other words, that outside the imaged area, there are infinitely many cyclic copies of the same source area, each contributing light. This introduces a small but noticeable amount of additional noise, which can be seen when we compare numerical simulations using the two methods.

\subsection{Deconvolution and the SNR}
\label{sec:SNR}

It is known that deconvolution disproportionately amplifies noise compared to the useful signal. To achieve a deconvolved image with a sufficiently high SNR, it is necessary to be able to estimate reliably this ``deconvolution penalty''.

In \cite{Turyshev-Toth:2020-extend}, we obtained just such an estimate by modeling the diagonal-dominated deconvolution matrix in the form (with appropriate normalization):
\begin{equation}
C_{ij}\to\tilde{C}_{ij}=\frac{4}{\pi\alpha d}\big(\mu\delta_{ij}+\nu U_{ij}\big),
\end{equation}
where $\delta_{ij}$ is the identity matrix, $U_{ij}$ is the ``everywhere-one'' matrix (i.e., $U_{ij}=1$ for all $i,j$), $\mu=1-\nu$ and $\nu\ll 1$ is given by
\begin{align}
\nu&=\frac{1}{N(N-1)}\Big(\sum_{i=1}^N\sum_{j=1}^NC_{ij}-\sum_{i=1}^NC_{ii}\Big)
=\frac{1}{AA'}\iint_{A'} d^2{\vec{x}'}\iint_{A} d^2\vec{x}\frac{d}{4|{\vec x}+\beta{\vec x}'|},
\end{align}
where we approximated the sum with a corresponding integral based on the approximate form of the averaged PSF, with $A'$ representing the source area in the source plane and $A=\beta^2A'$ its projection in the image plane. Using $\vec{x}''=-\beta\vec{x}'$, we can rewrite this expression as
\begin{align}
\nu=\frac{1}{A^2}\iint_{A}d^2{\vec{x}''}\iint_{A} d^2\vec{x}\frac{d}{4|{\vec x}-{\vec x}''|}.
\label{eq:AA}
\end{align}
To estimate $\nu$, first we assume that the integral does not depend on the choice of $\vec{x}''$, and use the central pixel of the image area as representative of all pixels. This is approximately true, since apart from pixels near the edge of the image, the majority of pixels are ``interior'' pixels, getting most blurred light from their neighboring pixels. This approach also accurately reflects the implied assumptions behind Fourier deconvolution, discussed at the end of the previous subsection. Under this assumption, the outer integral along with the normalization factor $1/A$ can be removed. The inner integral can be written in Cartesian coordinates as
\begin{align}
\nu=\frac{1}{N}\iint\limits_{x,y=-\sqrt{N}d/2}^{\sqrt{N}d/2} dxdy\frac{d}{4\sqrt{x^2+y^2}}\sim \frac{\ln(\sqrt{2}+1)}{\sqrt{N}}.
\label{eq:nu}
\end{align}

In this calculation, we estimated $\nu$ by assuming that pixels cover the image plane without either oversampling or undersampling the image. That is, given a square image of size $A=D^2$, $D=\sqrt{N}d$. If the image is over- or undersampled, i.e., if it is sampled at spatial intervals other than $d$, $D\ne\sqrt{N}d$, this scales the integrand in Eq.~(\ref{eq:nu}) by $\sqrt{N}d/D$:
\begin{align}
\nu\simeq \ln(\sqrt{2}+1)\frac{d}{D},
\label{eq:nu2}
\end{align}
where $d$ is the diameter of the telescope and $D$ is the characteristic linear size of the image formed on the image plane.

The inverse of the convolution matrix is approximated as
\begin{align}
C_{ij}^{-1} =
\frac{\pi\alpha d}{4}
\left[\frac{1}{\mu}\delta_{ij}-\frac{\nu}{\mu(\mu+\nu N)}U_{ij}\right]
 \simeq
\frac{\pi\alpha d}{4}
\left[(1+\nu)\delta_{ij}-\frac{1}{N}U_{ij}\right].
\end{align}

Defining the signal-to-noise ratio (SNR) as the ratio of the average signal level to the noise standard deviation, we introduce the deconvolution penalty: the relative amplification of noise at the expense of signal during deconvolution, which is given by
\begin{align}
\frac{{\rm SNR}_{\tt R}}{{\rm SNR}_{\tt C}}&=
\frac{\displaystyle\frac{1}{N}\sum_{i=1}^N\sum_{j=1}^N\, C^{-1}_{ij}}{\Big({\displaystyle\frac{1}{N}\sum_{i=1}^N\sum_{j=1}^N\,(C^{-1}_{ij})^2}\Big)^\frac{1}{2}~}\simeq\frac{\mu}{\nu N},
\label{eq:deconvpen0}
\end{align}
where ${\rm SNR}_{\tt R}$ and ${\rm SNR}_{\tt C}$ are, respectively, the signal-to-noise ratios of the recovered (deconvolved) image and the convolved imaging data collected by the telescope positioned in the image plane (see Fig~\ref{fig:SGL1}c).

One final point that needs to be considered is that when we use the averaged PSF given by (\ref{eq:thePSF}), its form was developed assuming a circular, not square, telescope aperture. The amount of light, thus the signal collected by such an aperture is scaled by $(\pi d^2/4)/d^2\sim\pi/4$. Putting it all together, using $(\ref{eq:nu2})$ in $(\ref{eq:deconvpen0})$ and $\mu\sim 1$, we obtain the final form of the deconvolution penalty:
\begin{align}
\frac{{\rm SNR}_{\tt R}}{{\rm SNR}_{\tt C}}\simeq\frac{\pi}{4\ln(\sqrt{2}+1)}\frac{D}{Nd}\sim 0.891\frac{D}{Nd}.
\label{eq:deconvpen}
\end{align}
When pixels fully cover the image plane, $D=\sqrt{N}d$, the deconvolution penalty is ${\rm SNR}_{\tt R}/{\rm SNR}_{\tt C}=0.891/\sqrt{N}$.

We note that while this is a useful estimation of the deconvolution penalty, confirmed through simulation that is discussed in the next section, the actual deconvolution penalty depends on the content of the image. We therefore find it useful to express the deconvolution penalty in a general form as
\begin{align}
\frac{{\rm SNR}_{\tt R}}{{\rm SNR}_{\tt C}}=\frac{a}{\sqrt{N}},
\end{align}
where the factor $a = \big(\pi/4 \ln(\sqrt{2}+1)\big) D/\sqrt{N}d\sim 0.891 \big(D/\sqrt{N}d\big)$ is driven by the sampling strategy.

\section{Results}
\label{sec:RESULT}

We developed a simulation of SGL image convolution and deconvolution in the C/C++ programming language. To compute the elliptic integral in (\ref{eq:thePSF}), the Boost C++ scientific library\footnote{\url{https://boost.org/}} was used.

The simulation is built around a simple but efficient implementation of the two-dimensional Cooley--Tukey fast Fourier transform algorithm\footnote{\url{https://en.wikipedia.org/wiki/Cooley-Tukey_FFT_algorithm}}. We are assuming an exoplanet image that is square in shape, with a pixel resolution that is an integral power of two. Thanks to the efficiency of the FFT algorithm, the code can readily process images as large as $8192\times 8192$ pixels on a desktop personal computer in a matter of minutes.

\begin{figure}
\includegraphics[width=0.4\linewidth]{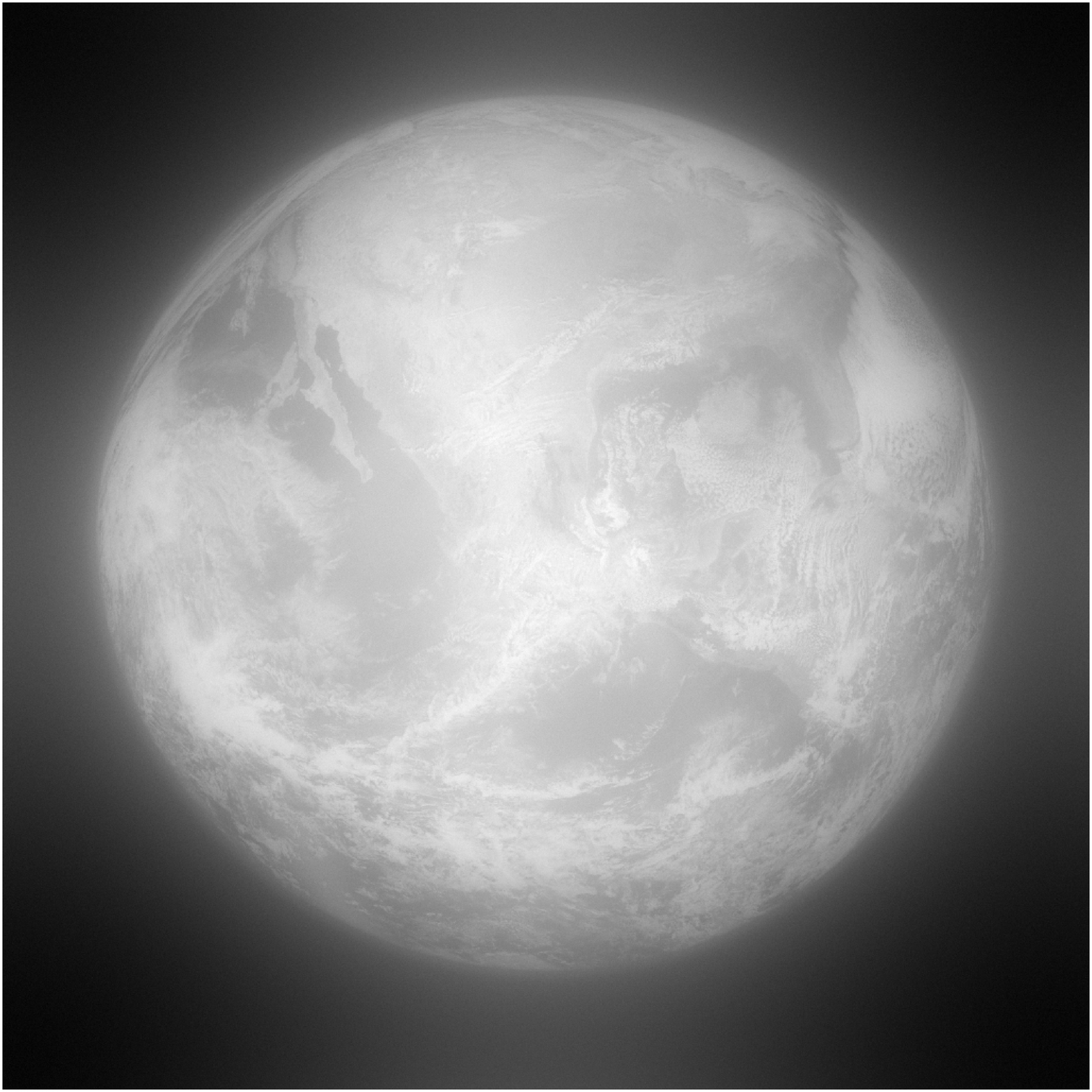}~\includegraphics[width=0.4\linewidth]{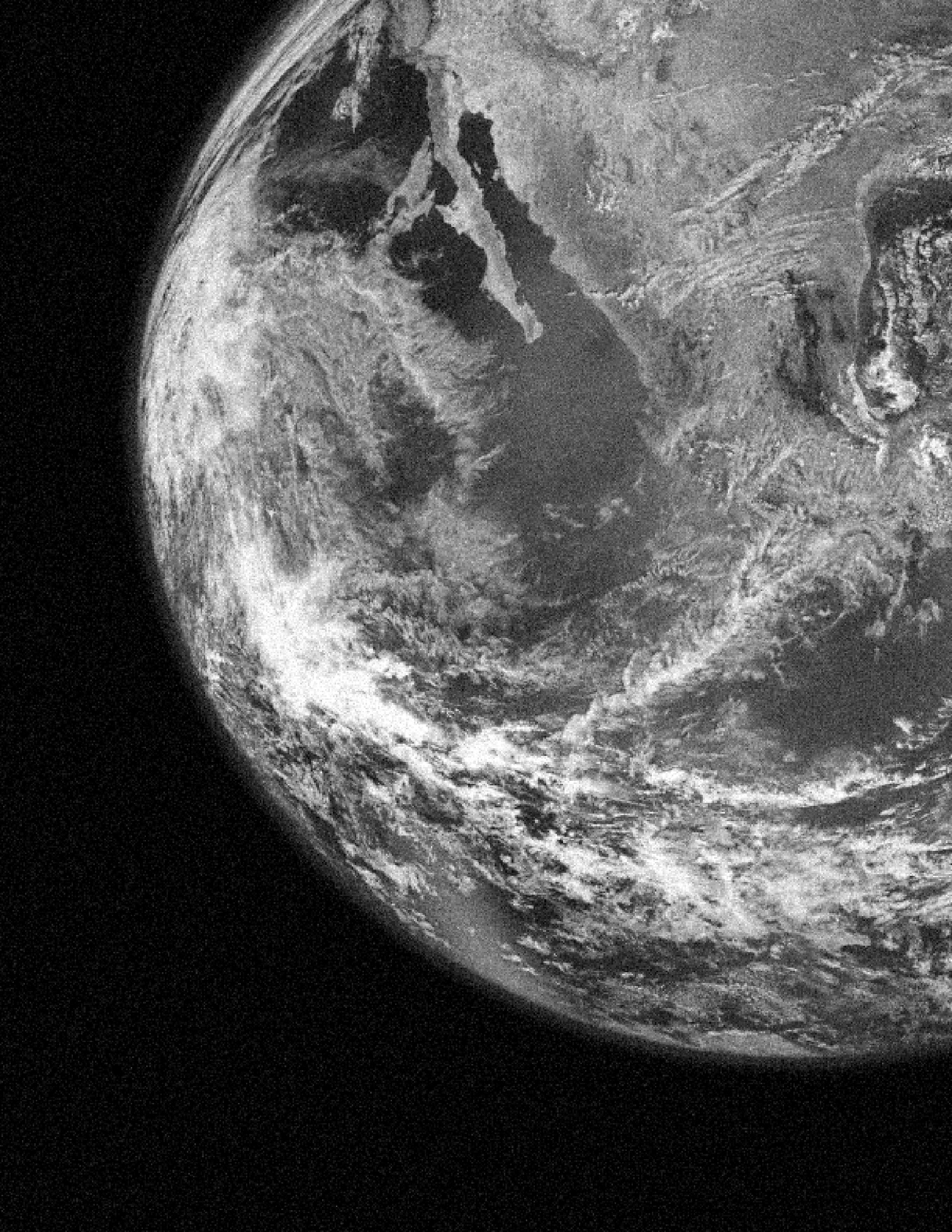}\\\vskip 1pt
\caption{\label{fig:PrCen}Image of a simulated Earth, at $1024\times 1024$ pixel resolution, at the distance of Proxima Centauri, at 1.3~pc, as projected by the SGL to an image plane at 650~AU from the Sun. Left: the convolved image with Gaussian noise added at ${\tt SNR}_{\tt C}=150$, corresponding to a cumulative integration time of $\lesssim 1$~year; right: the result of deconvolution with no noise filtering.}
\end{figure}

We used our simulation code on monochromatic images of the Earth as a stand-in for an exoplanet. Our primary goal at this point was to confirm, through simulation, the validity of our estimates of the effect of deconvolution on the SNR, and to also study possible techniques for noise suppression and improved image reconstruction.

First, we performed several simulations, using the Fourier method to create a convolved image, to which Gaussian noise was added at a predetermined SNR. Then, the image was deconvolved, again using the Fourier method, and the SNR was calculated by comparison with the original image. We used seven different monochrome images (two images of the Earth, a uniform white disk on a black background, a checkered pattern, a pure white and a pure black image and finally, an old television test pattern) and confirmed that across several resolutions (from $128\times 128$ to $1024\times 1024$ pixels) and with noise levels varying from ${\tt SNR}_{\tt C}=10$ to ${\tt SNR}_{\tt C}=10^5$ in $10,30,100,300,...$ increments, for fully sampled image planes ($D=\sqrt{N}d$) the deconvolution penalty was consistent with (\ref{eq:deconvpen}): ${\tt SNR}_{\tt R}/{\tt SNR}_{\tt C}=(0.878\pm 0.003)/\sqrt{N}$.

Our ultimate goal was to estimate the integration time required to obtain images of acceptable quality of remote, Earth-like targets, taking into account the non-removable stochastic noise due to the presence of the solar corona, through which the Einstein-ring is viewed. Given a pre-deconvolution value of ${\tt SNR}_{\rm C}$, we estimate the corresponding per-pixel integration time as \cite{Turyshev-Toth:2020-extend}:
\begin{align}
t_{\rm pix}=0.354\, \,{\tt SNR}^2_{\tt C}\, \bigg(1+0.79 \Big(\dfrac{650\,{\rm AU}}{z}\Big)^{5.1}
+0.05\Big(\dfrac{z}{650\,{\rm AU}}\Big)^{2.65}\bigg)
\Big(\frac{1\,{\rm m}}{d}\Big)^3\Big(\frac{z_0}{30\,{\rm pc}}\Big)^2\Big(\frac{650\,{\rm AU}}{z}\Big)^{3.4}~{\rm s},
\label{eq:tpix}
\end{align}
where $z_0$ is the distance to the Earth-like target and $z$ is the distance of the image plane from the Sun.

Our first objective was to investigate the case of a possible Earth-like planet at the distance of Proxima Centauri, $z_0=1.3$~pc. We can see how (\ref{eq:tpix}) can yield unrealistically long integration times when $D=\sqrt{N}d$; using $N=1024\times 1024$ and ${\tt SNR}_{\tt R}=5$ implies ${\tt SNR}_{\tt C}=5745$ and the corresponding per-pixel integration time for $z=1.3$~pc (the distance to Proxima Centauri) is over 40,000 seconds; for $1024\times 1024$ pixels in total, this amounts to 1340 years.

\begin{figure}
\includegraphics[width=0.4\linewidth]{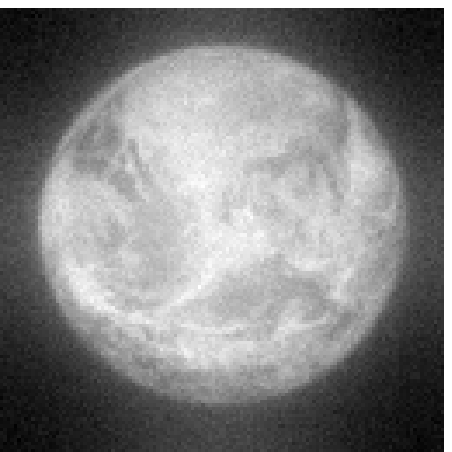}~\includegraphics[width=0.4\linewidth]{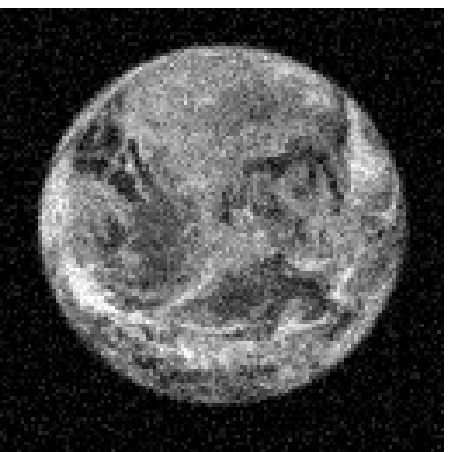}\\\vskip 1pt
\caption{\label{fig:z128}Simulated Earth at 30~pc, imaged at $128\times 128$ pixels at an image plane at 650~AU from the Sun. Left: the convolved image with Gaussian noise added at ${\tt SNR}_{\tt C}=50$, corresponding to a cumulative integration time of $\lesssim 1$~year; right: the result of deconvolution with no noise filtering.}
\end{figure}

However, this calculation fails to take into account the projected size of an Earth-like exoplanet image at $z_0=1.3$~pc, with the image plane at $z=650$~AU: the image is approximately $D=30$~km on one side. Thus, given a telescope aperture of $d=1$~m and $N=1024\times 1024$, we have $D/\sqrt{N}d=29.3$ and ${\tt SNR}_{\tt C}\lesssim 200$ is sufficient to achieve ${\tt SNR}_{\tt R}\sim 5$. This dramatically reduces the required integration time, to a mere 49 seconds per pixel, or a cumulative total of $\sim 1.6$ years for a $1024\times 1024$ pixel image.

Our next simulation was aimed at confirming this result, using a monochrome Earth image of $1024\times 1024$ pixels. After Fourier-convolution, noise at ${\tt SNR}_{\tt C}=150$ was added to the image and then the image was deconvolved using the parameters $D=30$~km, $d=1$~m. Confirming our calculations, the resulting, deconvolved image had ${\tt SNR}_{\rm R}=4.72$. This result, shown in Fig.~\ref{fig:PrCen}, is achievable with less than 1 year of cumulative integration time.

Next, we looked at the possibility of imaging more distant targets. As anticipated, the combined effects of reduced luminosity and smaller image area take their toll: the required integration time increases dramatically. The projected image size of an Earth-like exoplanet at 30~pc is $D\sim 1.3$~km. To obtain ${\tt SNR}_{\tt R}=5$ at $N=1024\times 1024$ requires a pre-deconvolution ${\tt SNR}_{\tt C}=4526$. The corresponding per-pixel integration time is $t_{\rm pix}=1.33\times 10^7$ seconds, which corresponds to nearly half a million years of cumulative integration time for a megapixel image, which is clearly impractical. However, even a modest reduction in resolution can greatly improve the chances of obtaining a usable image within a reasonable timeframe.

Specifically, if we aim at obtaining an image with $N=128\times 128$ pixels in the $1300\times 1300$~m${}^2$ image area, the situation changes dramatically. An image with ${\tt SNR}_{\tt C}=50$ can be obtained with a cumulative integration time of $\sim 0.85$~years. This corresponds to a post-deconvolution ${\tt SNR}_{\tt R}=3.5$, which is tolerable. Our actual simulation, shown in Fig.~\ref{fig:z128}, yielded a slightly better value of ${\tt SNR}_{\tt R}=4.1$.

Finally, we looked at the possibility of obtaining usable higher-resolution images by employing noise filtering. When a significant amount of noise is present, we found the Wiener deconvolution filter in Fourier space\footnote{\url{https://en.wikipedia.org/wiki/Wiener_deconvolution}} particularly effective. We implemented this filter with a single tunable parameter $K$, modifying (\ref{eq:deconv}):
\begin{equation}
\hat{O}(-\beta{\bf f})\propto\frac{\hat{I}({\bf f})}{\widehat{\tt PSF}({\bf f})}\cdot\frac{|\widehat{\tt PSF}({\bf f})|^2}{|\widehat{\tt PSF}({\bf f})|^2+K}.
\end{equation}
In all cases that we investigated, the parameter $K$ was hand-optimized to achieve a result with maximum post-deconvolution SNR.

\begin{figure}
\includegraphics[width=0.32\linewidth]{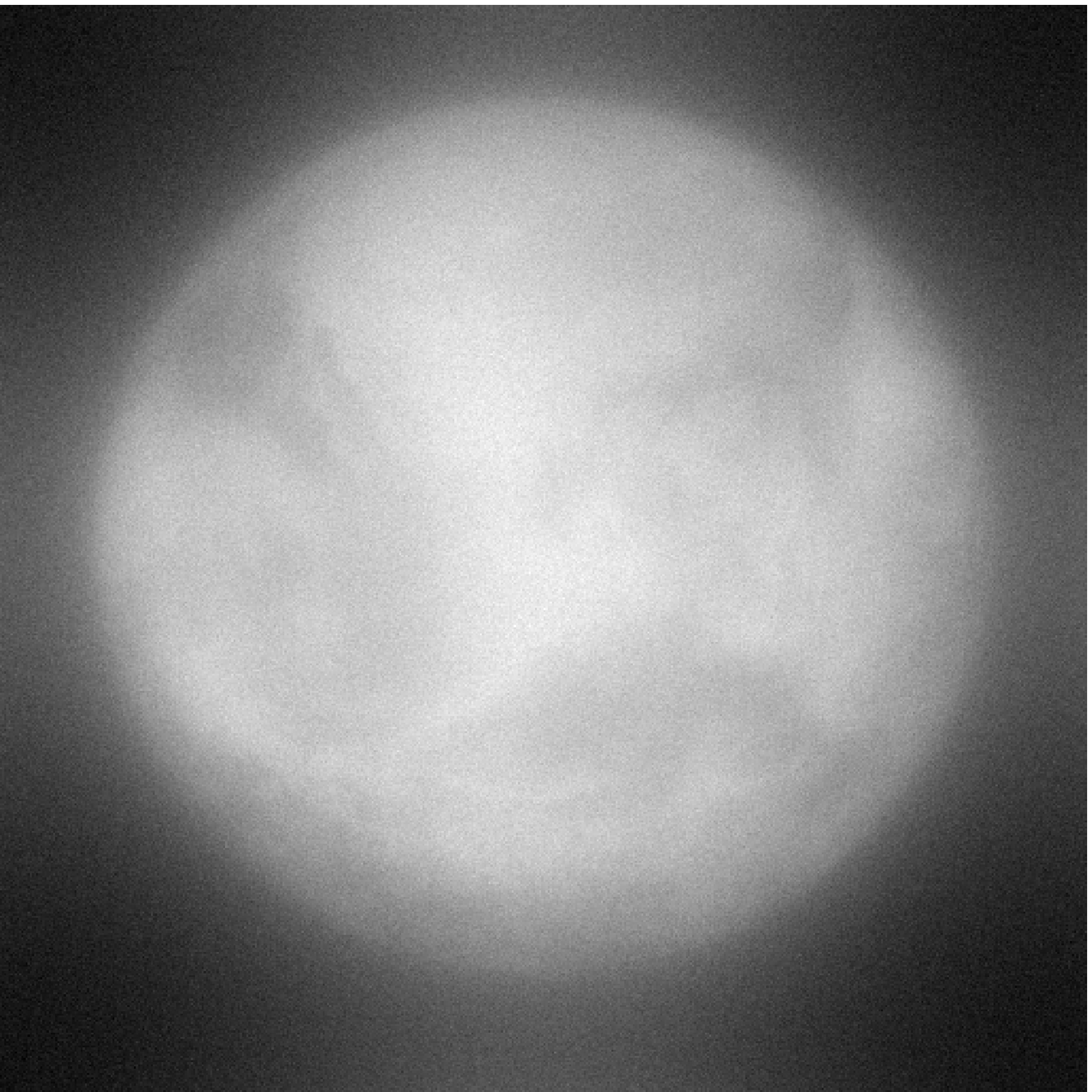}~\includegraphics[width=0.32\linewidth]{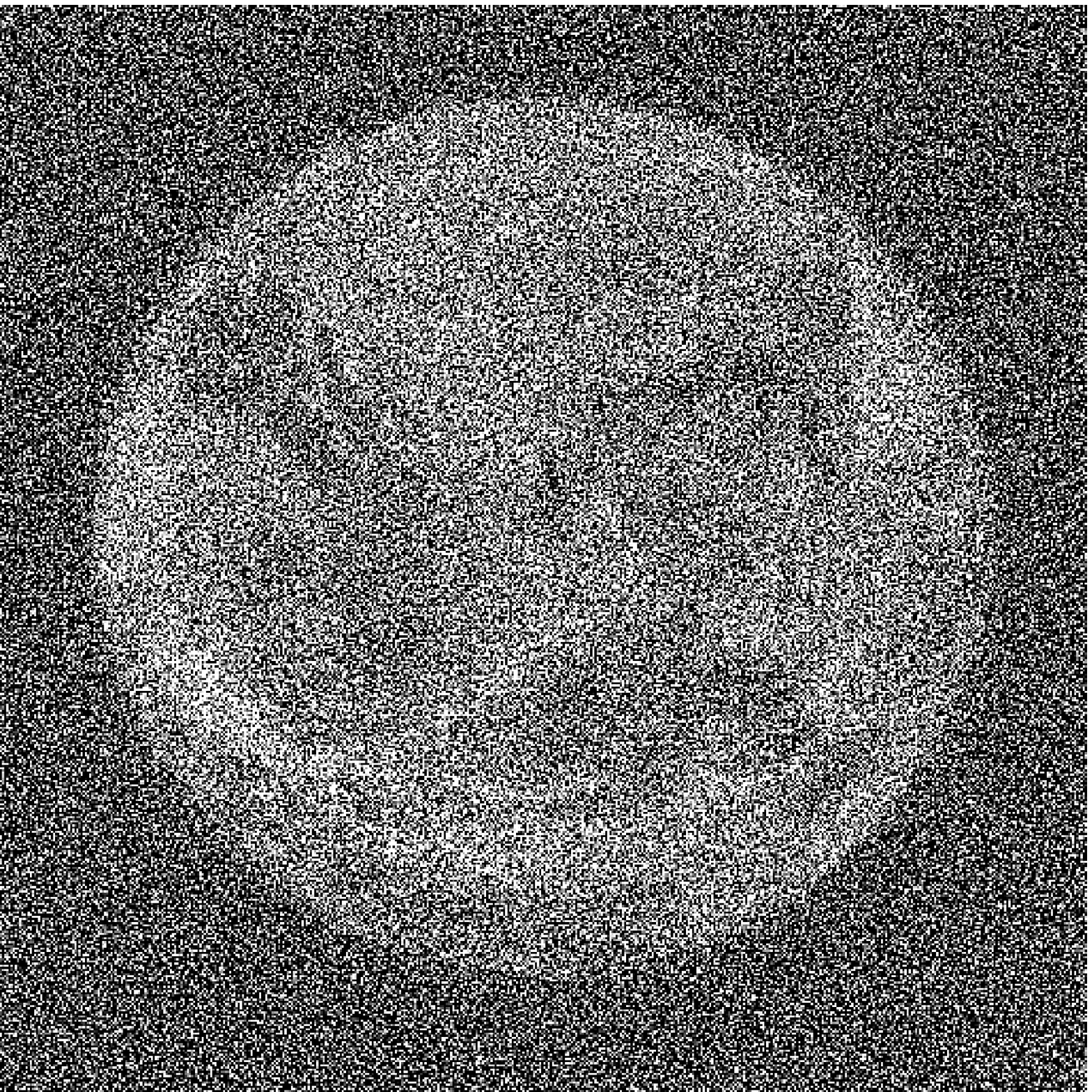}~\includegraphics[width=0.32\linewidth]{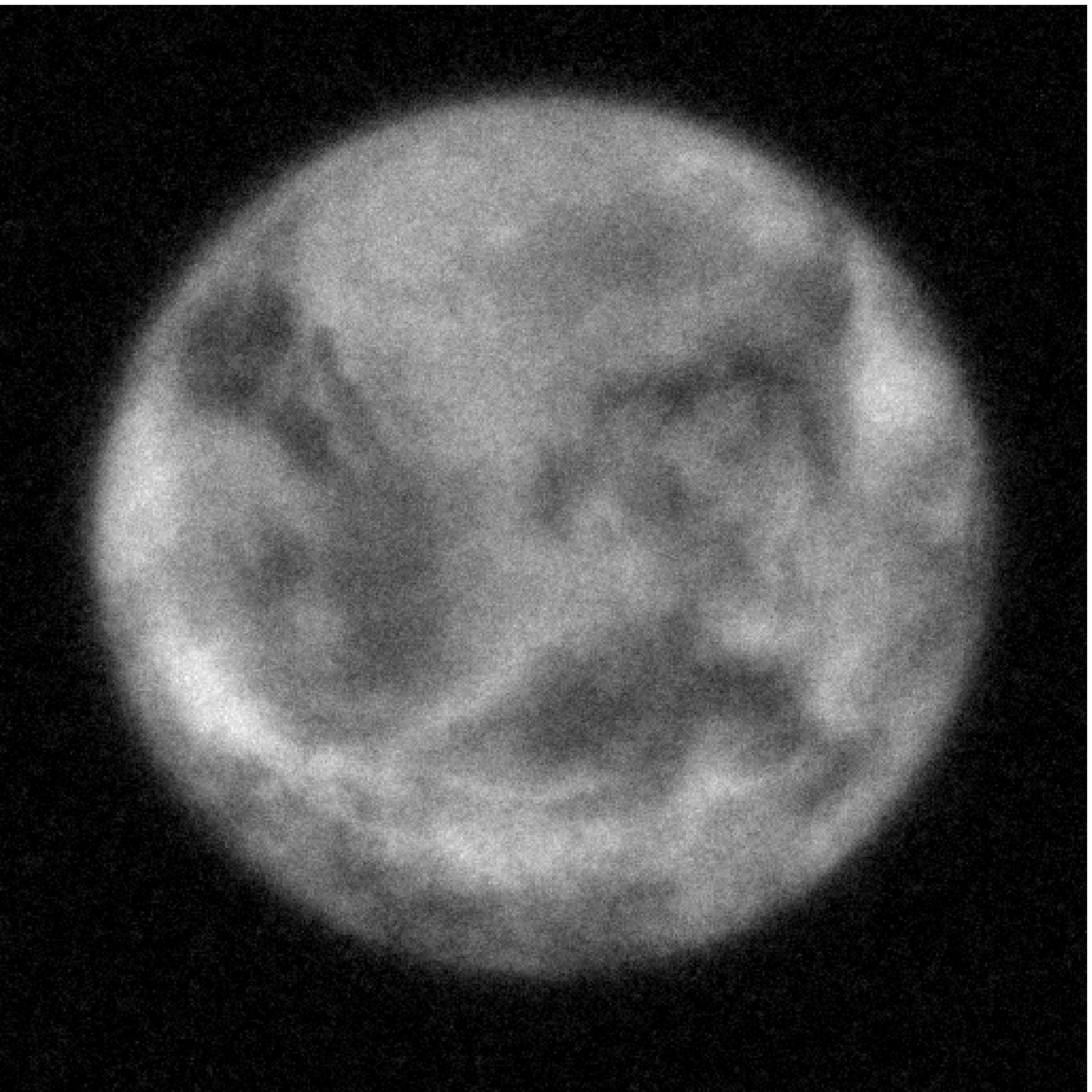}
\\\vskip 1pt
\caption{\label{fig:z512}Simulated Earth at 30~pc, imaged at $512\times 512$ pixels at an image plane at 650~AU from the Sun using a larger, 2-meter telescope. Left: the convolved image with Gaussian noise added at ${\tt SNR}_{\tt C}=50$, corresponding to a cumulative integration time of $\sim 1.7$~years; center: the result of deconvolution with no noise filtering; right: deconvolution using a tuned Wiener-filter.}
\end{figure}

As a specific case, we considered the same exo-Earth as before, at $z_0=30$~pc, but imaged with a $d=2$~m aperture telescope, at $N=512\times 512$ pixels of resolution. An image with a pre-deconvolution ${\tt SNR}_{\tt C}=50$ can be obtained in $\sim 1.7$ years. Deconvolution without noise reduction yields a very noisy image at ${\tt SNR}_{\tt R}=0.26$, with the outlines of the planet barely visible. Applying a tuned Wiener-filter, however, improves this to ${\tt SNR}_{\tt R}=3.1$ (Fig.~\ref{fig:z512}).

\section{Discussion and outlook}
\label{sec:CONCL}

Our initial experiments with our recently built image deconvolution code for the SGL offer useful insight.

The SGL projects an image of distant targets to an image plane located at $z>550$~AU from the Sun on the side opposite to the direction of the target. For a typical target, an exoplanet located $z_0<100$~light years from the Earth, the image area is measured in square kilometers. This projected image is sampled by a meter-class telescope that traverses the image plane while measuring the varying intensity of the Einstein-ring that forms around the Sun.

The SGL is an imperfect lens, characterized by spherical aberration. The result is a blurred projection, a convolved image. As the mathematical properties of the SGL are well-understood, it is possible to reconstruct, or deconvolve, the original from this blurred image. A major goal of our simulation was to improve our understanding of the deconvolution process and, in particular, its effect on the SNR of the resulting image.

The simulation confirmed the predicted drop in SNR, the ``deconvolution penalty'', characterized by (\ref{eq:deconvpen}). The penalty is proportional to the square root of the total number of image pixels (i.e., proportional to the linear image pixel density) but it is inversely proportional to the rate at which the image is undersampled. Consequently, choosing the rate at which the image plane is sampled plays a major role in our ability to collect enough information for good quality image reconstruction inside reasonable timeframes, consistent with realistic mission concepts to the SGL's focal region \cite{Turyshev-etal:2018,Turyshev-etal:2020}.

We found that realistic imaging scenarios can yield very high quality, megapixel resolution images of an Earth-like planet in a nearby solar system such as Proxima Centauri. Imaging planets in more distant solar systems is also possible at reduced resolution. At the extreme range that we considered, 30~pc ($\sim 100$~light years), a good quality image of an Earth-like planet can still be captured using a cumulative integration time of less than one year at $128\times 128$ pixels of resolution.

Additionally, it is possible to employ tailored noise reduction or noise suppression methods, such as the use of a Wiener-filter as part of the Fourier method of deconvolution. Such methods can further improve the SNR of the deconvolved image at the cost of a modest reduction in image resolution.

In this analysis, we assumed that all sources of non-stochastic noise can be estimated and their contributions can be removed from the signal, leaving only the stochastic component. In particular, being able to measure contributions from the solar corona reliably will be a significant challenge.

For this analysis, we considered a PSF that is averaged by the meter-class aperture of an observing telescope that is used to measure the overall intensity of the Einstein-ring around the Sun. This averaged PSF has no wavelength dependence. We have yet to investigate the possibility of using the SGL for spectral analysis and the impact of narrowband filters on the SNR.

In addition, we are yet to incorporate in the analysis the fact that the Sun's gravitational field is not truly spherically symmetric. Even small deviations from spherical symmetry (expected because of the oblateness and rotation of the Sun) can result in significant modification of the PSF, making it directionally dependent.

Furthermore, we have only considered targets that are stationary and fully illuminated. Obviously, a real exoplane will have varying illumination. Its appearance may also change due to planetary rotation, changes in cloud cover, or even surface changes such as those due to seasonal vegetation.

Finally, our current work assumed using only one imaging telescope. Clearly, using several instruments for imaging will improve the temporal sampling of imaging data. Such a capability will  allow for improvements in the understanding of temporally varying processes.

These topics are the subject of on-going study. Results, once available, will be reported elsewhere.

\begin{acknowledgments}
This work in part was performed at the Jet Propulsion Laboratory, California Institute of Technology, under a contract with the National Aeronautics and Space Administration.
VTT acknowledges the generous support of David Silver, Plamen Vasilev and other Patreon patrons.
\end{acknowledgments}


\end{document}